\documentclass[preprint,12pt,authoryear]{elsarticle}

\usepackage{amssymb}

\usepackage[table]{xcolor}

\usepackage{lineno}
\usepackage{hyperref}

\journal{Materials Chemistry and Physics}

\begin{document}

\begin{frontmatter}



\title{ Evaluation of Mater Bi and Polylactic Acid as materials for biodegradable innovative mini-radiosondes to track small scale fluctuations within clouds }

\address[label1]{Politecnico di Torino, Dipartimento di Scienza Applicata e Tecnologia, Corso Duca degli Abruzzi 24, 10129 Torino, I}
\address[label2]{Smart Materials, Istituto Italiano di Tecnologia, Via Morego 30, 16163 Genova, I}
\address[label3]{Istituto Nazionale di Ricerca Metrologica, Str. delle Cacce, 91, 10135 Torino, I}

\author[label1]{Tessa C. Basso}
\author[label2]{Giovanni Perotto}
\author[label3]{Chiara Musacchio}
\author[label3]{Andrea Merlone}
\author[label2]{Athanassia Athanassiou}
\author[label1]{Daniela Tordella\corref{cor1}\fnref{1}}

\fntext[1]{Corresponding author}

\address{}

\begin{abstract}
Turbulence plays an important part in determining the chemical and physical processes, on both the micro- and macro-scales, whereby clouds are formed and behave. However, exactly how these are linked together and how turbulence impacts each of these processes is not yet fully understood. This is partly due to a lack of in-situ small scale fluctuation measurements due to a limitation in the available technology. It is in this context that the radiosondes, for which the material characterisation is presented in this paper, are being developed to generate a Lagrangian set of data which can be used to improve the ever-expanding knowledge of atmospheric processes and, in particular, the understanding of the interaction between turbulence and micro-physical phenomenologies inside clouds (www.complete-h2020network.eu). Specifically, the materials developed for the balloons are discussed in further detail within this paper. Mater Bi and polylactic acid are the two common biodegradable thermoplastics that were used initially to make the balloons. To tailor their properties, the balloons were then coated with carnauba wax blended with either pine resin or SiO$_2$ nanoparticles. The properties such as hydrophobicity, toughness, elasticity and helium gas permeability are investigated and improved in order to keep the density of the radiosondes as constant as possible for a couple of hours. This will allow them to float inside and outside clouds on an isopycnic surface, to measure various properties such as velocity, temperature, pressure and humidity by means of solid state sensors and to transmit them to receivers on Earth. Tests have been made under a rigorous metrological approach comparing the 6 new materials with two reference balloon materials, latex and mylar. It was found that Mater Bi with the two carnaubua wax coatings is the most suited though its roughness and water vapour permeability should be improved.

\end{abstract}


\end{frontmatter}


\section{Introduction}
\label{}
Clouds play a key role in climate change and climate sensitivity as they are responsible for approximately 31 \% of rain on the planet \cite{lau2003warm} and their albedo, which is dependent on water droplet concentration, influences the Earth's energy balance \cite{boucher2013clouds, bodenschatz2010can}. They are a multi-scale natural phenomena extending on a wide range of chemical and physical processes rendering them one of the largest sources of uncertainty in weather prediction and climate modelling. It is necessary to establish a connection from aerosol and particle microphysics \cite{grabowski2013growth,prabhakaran2017can, devenish2012droplet} to macro-scale turbulent dynamics such as entrainment \cite{mellado2017cloud, de2013entrainment}. However, though there are many instruments in use to study the atmosphere, from satellites \cite{radkevich2008scaling} to dropsondes \cite{bertoldo2016hail, tsai2011floating} to big weather balloons \cite{pommereau2015observations, businger2006scientific}, there is a lack of in-situ measurements along Lagrangian trajectories to track fluctuations of the physically relevant quantities on scales ranging from a few centimetres to a few kilometres.

Within the Horizon 2020 Innovative Training Network Cloud-MicroPhysics-Turbulence-Telemetry (MSCA-ITN-COMPLETE) project \cite{complete2016webpage} a new kind of small, green, ultralight radiosonde capable of floating in warm clouds is being developed \cite{basso2017disposable}.  They will float on an isopycnic, constant density, surface at a chosen altitude between 1-3 km for a time range spanning from the inner turbulence time scale (a few minutes) to the extension of cloud lifetime (a few days). These sondes are conceived to measure fluctuations in temperature, pressure, humidity, acceleration, and aerosol concentration. They will collect data using specifically chosen solid state sensors and will temporarily store them on a flash memory before transmitting them to receivers on Earth through the LoRa protocol \cite{wixted2016evaluation, paredes2018ultra}. This will provide an insight into Lagrangian fluctuations inside warm clouds over land and alpine environments \cite{businger2006scientific, businger1996balloons}.

It should be noted, that this specific kind of radiosonde must be as small as possible (about 30 $cm$ in diameter) because in small scale cloud Lagrangian experiments the radiosonde must not only have a minimal volume compared to its trajectory length, but, most importantly, must be apt to passively follow cloud fluctuations. Consequently, their  weight is limited to a global weight of approximately 20 $g$. However, to float on an isopycnic surface, the weight and volume of the balloon must remain relatively unchanged throughout the duration of the flight. Hence, the balloon material should not be too elastic to keep the volume constant, it should be hydrophobic as clouds contain water droplets and any that can be absorbed by or adhere to the balloon surface will change its weight, and finally, they need to be impermeable to helium, He. Furthermore, the electronics and sensors used must be chosen carefully and, for example, the use of GPS is not envisaged. Consequently, the radiosondes will not be recovered once their flight is over. It becomes clear that to limit the environmental impact of the radiosondes, they must be as biodegradable as possible as well as cheap. 

The first untethered balloons were made of an elastic material \cite{du2002invention} but for the balloons developed within this project the use of fixed-volume balloons is necessary. The most common materials used are polyester, fabric or a polyethylene composites. Good examples of these are latex, a rubber material and mylar, a polyethylene terephthalate polymer containing aluminium. The latter is lightweight, has a low gas permeability and is incredibly tough. It falls into the category of thermoplastic polymers which makes its processing easy. However, it is not biodegradable and contains metal that interferes with the transmitted signals when the data is sent to receivers on Earth. Therefore, the materials considered for the replacement of these two references are Mater Bi and Polylactic Acid (PLA); thermoplastic polymers partly derived from natural sources \cite{tran2017starch, la2018fully}. They are both abundant and inexpensive polymers which are easy to process and tune to obtain tailored properties by either adding compounds during synthesis or later by means of layers which can be sprayed on \cite{bayer2014direct, heredia2017all}. Additionally, they are both degraded by various types of bacteria making them promising candidates for the present application \cite{li1995degradable}.

The focus of this paper is on the biodegradable material development of the exterior enclosure of the radiosondes described above. The synthesis as well as the characterisation of these materials will be discussed. The tests focused on characterising their mechanical strength as well as their surface properties to compare the materials to the two references. The selection of the most appropriate materials for the purpose is done based on which can best imitate the desired properties from the references. Furthermore, tests were conducted to characterise the materials' specific features such as: their insulation from humidity, and temperature inertia inside the balloon, to protect the electronics; and permeability, to limit weight increase during flight. Measurements have been performed at the Italian National Metrology Institute (INRiM), using equipment specifically developed for applied thermodynamics for meteorology and climate, in the framework of the MeteoMet project \cite{merlone2018meteomet2, merlone2015meteomet, musacchio2015arctic}. The procedures adopted for this measurement process were defined under a rigorous metrological approach, involving traceable standards of temperature and humidity. They were performed in a specific climate chamber highly effective in terms of stability and time required to change between humidity and temperature set points and possessing a high resolution read out system, such as resistance thermometers for reading PRT100. The uncertainty in all the measurements was evaluated in order to allow a better comparability among the different materials tested. The same equipment is planned to be involved in the further characterisation of the sensors that will be embedded in the radiosonde system, inside the balloons.

\section{Experimental}
\label{}
\subsection{Materials}
As stated above, the reference materials for this paper are latex and mylar. These were store bought in a local store specialised for balloons. The Mater Bi used in this study are store bought Mater Bi biodegradable bin bags.  They are very thin, 30 $\mu$m and have a density of 1.3  $g\hspace{0.1cm}cm^{-3}$ \cite{bastioli1998properties}. The PLA used in this study originate as pellets of grade 4043D and were bought from Sigma-Aldrich. They have a density similar to that of Mater Bi, with a value of 1.24 $g\hspace{0.1cm}cm^{-3}$ \cite{garlotta2001literature}. The two solvents used within this study were chloroform ($\geq 99.5 \%$) and acetone ($\geq 97 \%$), both bought from Sigma-Aldrich. The coatings were made using carnauba wax (Cwax) bought from Sigma-Aldrich, a common vegetable wax derived from the leaves of a Copernicia palm, blended with two different substances; pine resin sourced from a local supplier and silica nanoparticles (SiO$_2$ NPs) of 25 $nm$ size from Alfa Aesar.

\subsection{Processing Materials}

In this paper, balloons of 14 cm diameter with thicknesses ranging between 30 - 60 $\mu$m were fabricated with Mylar, PLA and Mater Bi.
A flowchart of the procedure to produce balloons is depicted in fig. \ref{fig:balloon process}. As shown, if the polymer is acquired in the form of pellets, as is the case for PLA, the first step in the process is to form films. Hence, 5 $g$ of PLA pellets were dissolved in 50 $ml$ of chloroform and stirred with a magnet stirrer at room temperature, as shown in fig. \ref{fig:balloon process}. The resulting, relatively viscous solution is either drop casted in a non-stick Teflon tray or bar coated onto a Teflon sheet. Once dry, the sheets could be pealed off of the Teflon substrate. When the PLA film was half way through evaporating, it was folded onto itself with a Teflon sheet cutout in the shape of a balloon in between to take advantage of the ability of PLA to bond to itself and hence create a preamture seal. This sped up the process as the melting point of PLA is between 145-160 $^{\circ}$C.

Mater Bi films could be made in the same way, however, in this study, the material was bought as sheets and it was therefore not necessary. It also has a low melting point of 60 $^{\circ}$C, which makes it very easy to thermally seal. Likewise, mylar was bought as sheets and sealed into the required shapes and tested. The latex balloons were bought as balloons and hence did not undergo any synthesis or processing.

Two sheets were then placed together in a carver press, separated by a Teflon sheet cutout in the shape of the balloon shown in fig. \ref{fig:balloon process}. A load of 0.1 metric tons was applied at the melting temperature of the polymer and subsequently cooled to ensure a complete seal of the shape and to homogenise the thickness of the sheets. Once they were cooled, the samples were cut to size; ready for inflation with He and immediately sealed to avoid any losses. 

\subsection{Coating of polymers}

The hydrophobic properties were achieved by means of spray coating. Four solutions were prepared: 1. 1 $g$ carnauba wax dissolved in 50 $ml$ acetone; 2. 0.5 $g$ carnauba wax, 0.5 $g$ pine resin in 50 $ml$ acetone; 3. 0.98 $g$ carnauba wax dissolved in 40 $ml$ acetone to which was added 0.02 $g$ SiO$_2$ NPs in 10 $ml$ acetone; 4. 0.49 $g$ carnauba wax, 0.49 $g$ pine resin dissolved in 40 $ml$ acetone to which was added 0.02 $g$ SiO$_2$ NPs in 10 $ml$ acetone. The SiO$_2$ NPs were prepared by ultrasonication (for 30 minutes) separately. After mixing them together, 5 mL of the solutions were sprayed, following the procedure described in reference \cite{wang2016superhydrophobic}.

\subsection{Contact Angle}

The contact angle, created when a  5 $\mu l$ drop of water is deposited on the surface of the films, was measured using the Contact Angle System OCA from Dataphysics. The volume of the deposited drop was set to allow a good measurement of the surface properties to be taken without interference of other phenomena such as gravity which can be important if the volume of the drop of water is too big. However, if the material is superhydophobic and the drop would not leave the nozzle of the needle, the volume was set to 12 $\mu l$. Both pristine and treated surfaces were tested. 

\subsection{Tensile Strength}

The mechanical characterisation of the samples was done using an INSTRON 3365 uniaxial testing machine. The samples were cut into the standard dog bone shape, 25 mm in length, 3.98 mm in width with the thickness being that of the material, and the straining rate was set at 1 $mm\,min^{-1}$. 

\subsection{Helium Permeability}

With a density of 0.164 $kg\hspace{0.1cm}m^{-3}$  , He is much lighter than air, whose density is 1.225 $kg\hspace{0.1cm}m^{-3}$ . Hence, a balloon that is inflated with He will be lighter, and consequently weigh less until the volume of He will be such that the balloon will begin to float. This is the principle upon which the test has been designed. By measuring the increase in weight, after inflating the balloon with He, over time, the ability for the material to keep He under pressure within it can roughly be evaluated and compared with other materials. To do this, the balloons were weighed before the measurement, inflated to maximum capacity and immediately thermally sealed with a hot electric rod. It was immediately put on an analytical weighing scale to isolate the balloon from any air currents. The weight was recorded at regular intervals, with increased frequency at the beginning of the measurement, for a total duration of 200 hours.

\subsection{Temperature and humidity measurements in climate chambers}

Measurements were performed in the Applied Thermodynamics laboratory of INRiM. The instrumentation involved in the characterization of the mini-radiosonde materials is detailed bellow. The Climate Chamber Kambic KK190 CHLT MeteoCal allows the control of temperature in the range of -40 $^{\circ}$C to 180 $^{\circ}$C and relative humidity from 10 \% to 98 \%. Chamber temperature stability during measurements has been evaluated to be 0.02 $^{\circ}$C with temperature homogeneity in the measurement area of 0.05 $^{\circ}$C. Reference temperature sensors used are CalPower PT100, 4-wire platinum resistance thermometers with 3 mm diameter. Sensors are calibrated at INRiM and a Fluke 1594A Super-Thermometer is used for readings. The uncertainty contribution on temperature measurements due to reference sensor calibration is 0.01. The relative humidity (RH) is measured with a Delta Ohm Humidity and Temperature Probe model HD27.17TS and datalogger HD28.17T.DO. Calibration Uncertainty of the relative humidity sensor at 10 $^{\circ}$C is 1.6 \% RH with a resolution of 0.1 \% RH. Sensors has been selected to be the best compromise between accuracy, robustness, and response speed in order to evaluate the material behaviour in a temperature and humidity controlled environment during the testing.

The first test evaluated the effect of condensation of the temperature readings. To create condensation, the chamber was initially set to 40 $^{\circ}$C with a relative humidity of 10 $\%$, i.e. dry conditions. Once the values were stable, the chamber temperature was set to 10 $^{\circ}$C (keeping RH at 10 $\%$) and the resulting curve was recorded (dry cycle). Without opening the chamber, this cycle was repeated by setting the chamber to 40 $^{\circ}$C again but this time with a RH of 98 $\%$ and brought down to 10 $^{\circ}$C, 98 $\%$ RH (wet cycle). 

For the second test, at a constant temperature of 10 $^{\circ}$C, common in warm clouds  \cite{malinowski2013physics}, humidity was set from 10 $\%$ to 98 $\%$ and the relative humidity inside the balloon was monitored. The reference humidity was taken from the chamber sensor itself. After a drying procedure (40 $^{\circ}$C and 10 $\%$ for 1 hour), the chamber was set to an initial condition of 10 $^{\circ}$C and 10 $\%$ relative humidity. The chamber was then opened and the balloons were inflated with the ambient air within to try and obtain the same starting points both inside and outside of the balloon. The chamber was closed and left to stabilise again before setting the humidity to 98 $\%$ and monitoring the humidity within the balloon for 1 hour.

\section{Material characterisation: Results and Discussion}

The size required for the radiosondes to float at a given altitude was determined using Archimedes' principle for buoyancy \cite{yajima2009engineering}. The equation for the stable floating condition is 

\begin{equation} \label{archimedes}
    V_b = \frac{m_t}{\rho_a (1-\frac{M_g}{M_a})} 
\end{equation}

where $V_b$ is the volume of the balloon, $m_t$ is the total mass of the system, $\rho_a$ is the density of the atmosphere at that altitude, $M_g$ and $M_a$ are the molecular weights of the gas and air, respectively. 

Assuming a constant lapse rate, $\gamma$ of 6.5 $K\, km^{-1}$, the results of the calculation are shown in table \ref{atmosphere}. Given that the density the radiosonde must match the atmospheric density at a certain altitude, its volume can be calculated. Hence, both weight and volume should be kept constant.

\begin{table} [!h]
\rowcolors{2}{yellow!80!white!70}{yellow!90!white!20}
\begin{tabular}  { p{1.5cm} p{1.5cm} p{2.4cm} p{1.3cm} p{1.5cm} p{3cm}}

 \hline
  \multicolumn{4}{p{8cm}}{\textbf{Atmospheric Data}} &  \multicolumn{2}{p{5cm}}{\textbf{Balloon Dimensions}}  \\
\hline
   Z [m] & T [K] & P x$10^4$ [Pa] & $\rho$ [$\frac{kg}{m^3}$]  & V [$m^3$] & R [cm] \\
   
 \hline
0 & 288 & 10.0 & 1.22 & 0.019 & 16.5 \\
 \hline
500 & 285 & 9.5 & 1.17 & 0.020 & 16.8 \\
 \hline
750 & 283 & 9.3 & 1.13 & 0.020 & 16.9 \\
 \hline
1000 & 282 & 9.0 & 1.11 & 0.021 & 17.1 \\
 \hline
  1250 & 280 & 8.7 & 1.08 & 0.021 & 17.2 \\
 \hline
1500 & 278 & 8.5 & 1.06 & 0.022 & 17.7 \\
 \hline
2000 & 275 & 7.9 & 1.01 & 0.023 & 17.7 \\
\hline 
3000 & 269 & 7.0 & 0.90 & 0.026 & 18.3\\
\hline

\end{tabular}
 \caption{The various atmospheric values which determine the balloon size by equating the density of the atmosphere to that of the balloon (at a fixed total weight of 20 gr). Here, Z is the altitude, T is the temperature, P is the pressure, $\rho$ is the atmospheric density, V is the volume of the balloon while R is its radius. These values were calculated at a constant lapse rate $\gamma$ = 6.5 $K km^{-1}$ \cite{basso2017disposable}.}
 \label{atmosphere}
\end{table}
 
\begin{figure} 
    \centering
     \includegraphics[width=0.8\textwidth]{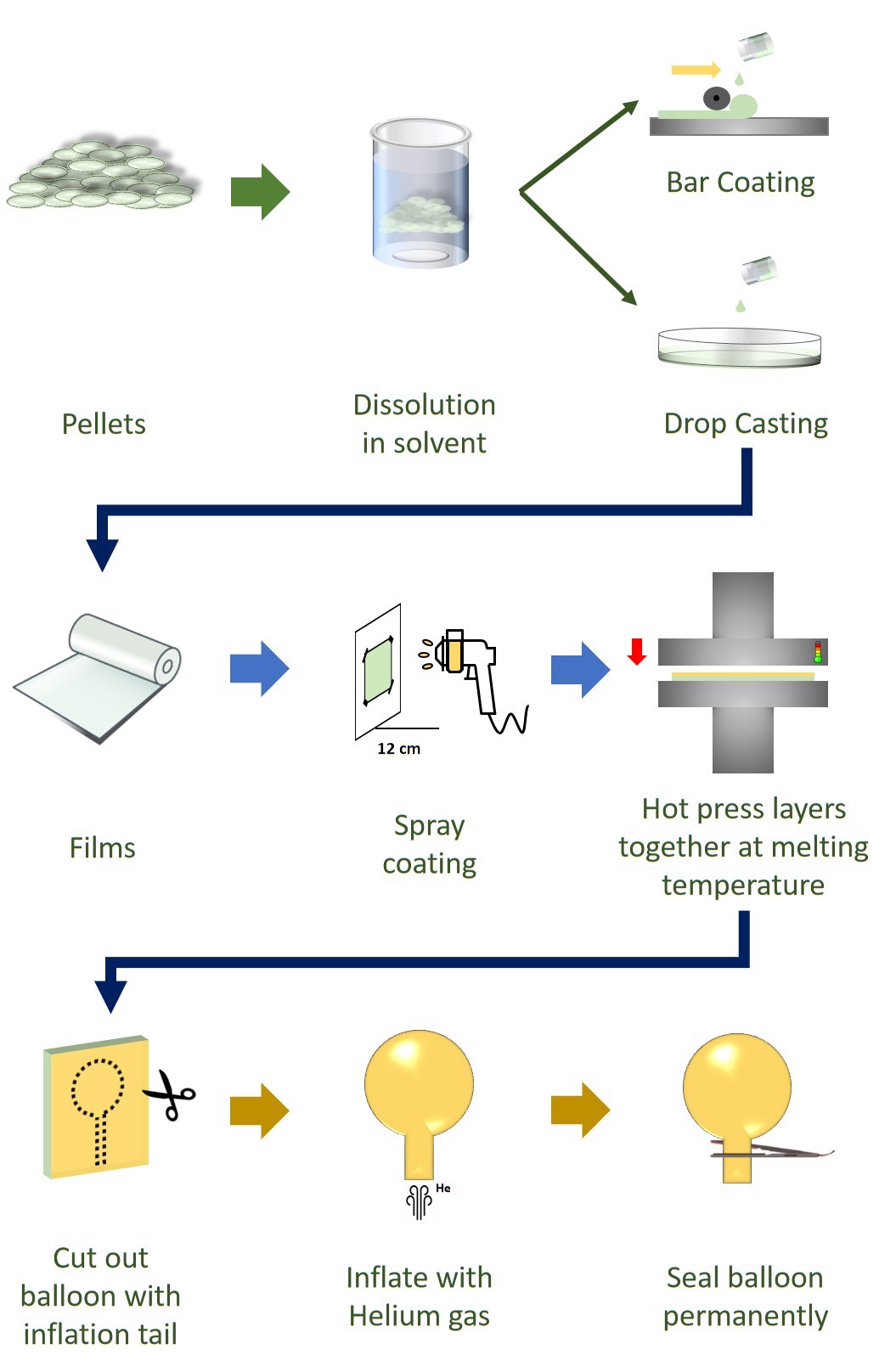}
    \caption{Flow chart of balloon making process. The first row is relevant for compounds that are acquired in the form of pellets and describes how the films used in this study are made. The second row is about the surface treatments that were done. If the material comes in the form of sheets, the procedure usually starts from the second row. The last steps are specific to making the final balloons.  }
    \label{fig:balloon process}
\end{figure}

\begin{figure}[h!]
    \centering
    \includegraphics[width=0.8\textwidth, height=6cm]{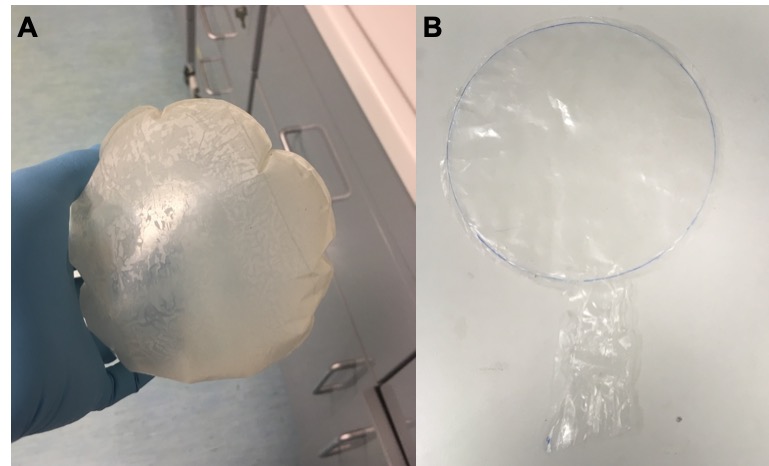}
    \caption{Pictures of the final balloon products where A) is an inflated Mater Bi balloon and B) PLA balloon before inflation.}
    \label{fig:balloon}
\end{figure}

The assembled process to make the balloons is seen in fig. \ref{fig:balloon process}. The solvent processing techniques, namely drop casting and bar coating, were chosen as they enable the thickness of the films to be somewhat controlled with a certain amount of precision as it mainly depends on the concentration of the solvent/polymer solution. Additionally, the techniques are easily scaled up to obtain larger size samples and produce films that are relatively homogeneous. Similarly, spray coating techniques were employed as they are very attractive for scaled up industrial applications because they don't require vacuum, unlike other vapour depositions \cite{girotto2009exploring}. Additionally, it is a versatile technique as the properties of the resulting film depend on the spray parameters such as distance from target, pressure, solution concentration and number of layers. As mentioned previously, blends of carnauba wax with pine resin or SiO$_2$ nanoparticles were tested to improve the surface properties of Mater Bi and PLA. These were chosen as resins are typically insoluble in water, derive from plants and have the ability to be converted into a polymer like substance while SiO$_2$ NPs are able to chemically bond to hydrophobic groups.

\begin{figure}[h!]
    \centering
    \includegraphics[width=1\textwidth]{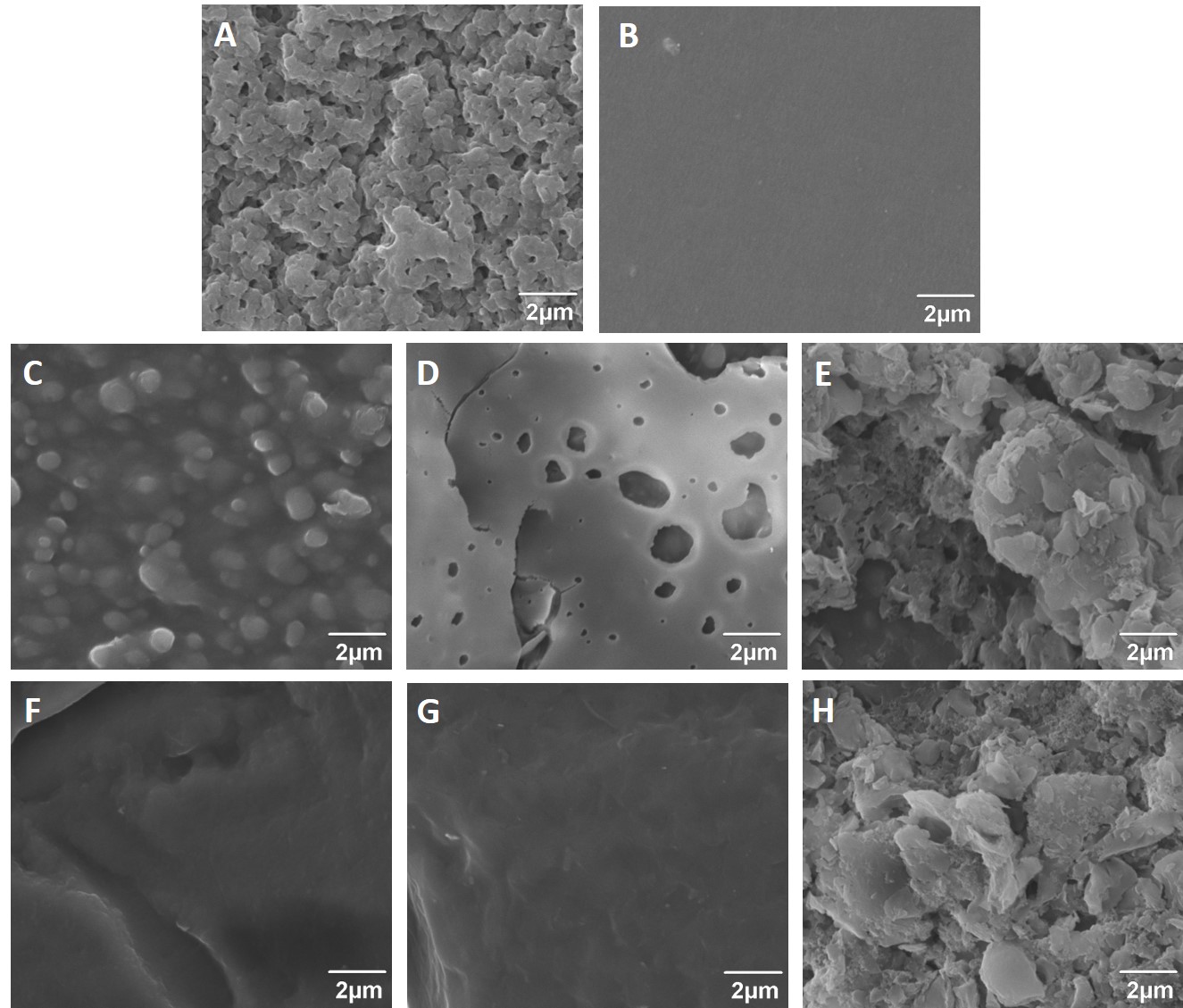}
    \caption{SEM images of all the materials at magnification x6000. A) Latex, B) Mylar, C) Mater Bi, D) Mater Bi + Carnauba Wax + Pine Resin, E) Mater Bi + Carnauba Wax + SiO$_2$ NPs, F) PLA, G) PLA + Carnauba Wax + Pine Resin, H) PLA + Carnauba Wax + SiO$_2$ NPs.}
    \label{fig:sem}
\end{figure}

The process depicted in fig. \ref{fig:balloon process} resulted in homogeneous balloons of similar size (14 $cm$) and thickness (between 30 $\mu$m - 60 $\mu$m) for which examples are shown in fig. \ref{fig:balloon}. Images of these were taken with a Scanning Electron Microscope, SEM, and reported in fig. \ref{fig:sem} at a magnification of x6000 and 10 kV.

\subsection{Tensile Strength}

As mentioned previously, the balloons are set to be fixed-volume because if they were allowed to expand indefinitely, they would do so until they popped \cite{du2002invention}. Additionally, by expanding, the overall density of the radiosonde would be altered, changing the altitude at which it floats. Hence, some flexibility is required for the balloon to bare any overpressure whilst maintaining a resistance to substantial expansion \cite{nakai2012high}.

\begin{table}[h!]
\footnotesize
{\rowcolors{2}{yellow!80!white!70}{yellow!90!white!20}
\begin{tabular}{ p{2.3cm}| p{1.8cm} p{1.8cm} p{1.8cm} p{1.8cm} p{1.8cm}}

 \hline
\textbf{Material} & \textbf{Young's} & 
\textbf{Tensile stress} & \textbf{Ultimate} & \textbf{Elongation} & \textbf{Energy}\\
& \textbf{Modulus} & \textbf{at yield} & \textbf{stress} & \textbf{at break} & \textbf{($J/m^3$)}\\
& \textbf{(MPa)} & \textbf{(MPa)} & \textbf{(MPa)} & \textbf{$(mm/mm)$} & \\
 \hline
 \textbf{Latex} & 4 $\pm$ 1 & 7 $\pm$ 1 & 27 $\pm$ 1 & 18 $\pm$ 0.3 & 125 $\pm$ 6\\
 \hline
 \textbf{Mylar} & 1185 $\pm$ 33 & 30 $\pm$ 1 & 55 $\pm$ 3 & 0.8 $\pm$ 0.1 & 35 $\pm$ 4\\
 \hline
 \textbf{Mater Bi} & 161 $\pm$ 13 & 8 $\pm$ 1 & 12 $\pm$ 1 & 3.2 $\pm$ 0.1 & 34 $\pm$ 2\\
 \hline
 \textbf{Mater Bi\hspace{0.05cm}+\hspace{0.05cm}Cwax\hspace{0.05cm}+\hspace{0.05cm}Pine Resin} & 121 $\pm$ 10 & 6 $\pm$ 0.2 & 10 $\pm$ 1 & 3.1 $\pm$ 0.7 & 26 $\pm$ 6\\
 \hline
  \textbf{Mater Bi\hspace{0.05cm}+\hspace{0.05cm}Cwax\hspace{0.05cm}+\hspace{0.05cm}SiO$_2$ NPs} & 108 $\pm$ 11 & 4 $\pm$ 1 & 7 $\pm$ 0.5 & 1.1 $\pm$ 0.1 & 8 $\pm$ 1\\
 \hline
\textbf{PLA} & 1578 $\pm$ 93 & 18 $\pm$ 2 & 24 $\pm$ 4 & 0.09 $\pm$ 0.05 & 4 $\pm$ 2\\
 \hline
 \textbf{PLA\hspace{0.05cm}+\hspace{0.05cm}Cwax\hspace{0.05cm}+\hspace{0.05cm}Pine Resin} & 925 $\pm$ 22 & 15 $\pm$ 1 & 17 $\pm$ 1 & 0.07 $\pm$ 0.04 & 6 $\pm$ 3\\
 \hline
 \textbf{PLA\hspace{0.05cm}+\hspace{0.05cm}Cwax\hspace{0.05cm}+ SiO$_2$ NPs} & 993 $\pm$ 112 & 14 $\pm$ 1 & 17 $\pm$ 1 & 0.06 $\pm$ 0.01 & 2 $\pm$ 1\\
 \hline
\end{tabular}
}
 \caption{The characteristic average values for the Young's Modulus, Tensile stress at yield, the ultimate stress that the material can bare before breaking, the elongation and the total absorbed energy by the material before breaking. These measurements were performed with an INSTRON machine at a deformation rate of 1 {$mm\hspace{0.1cm}min^{-1}$}. The value following the plus or minus sign represents the standard deviation of the parameter.}
 \label{strength}
\end{table}

\begin{figure}[h!]
    \centering
    \includegraphics[width=1\textwidth]{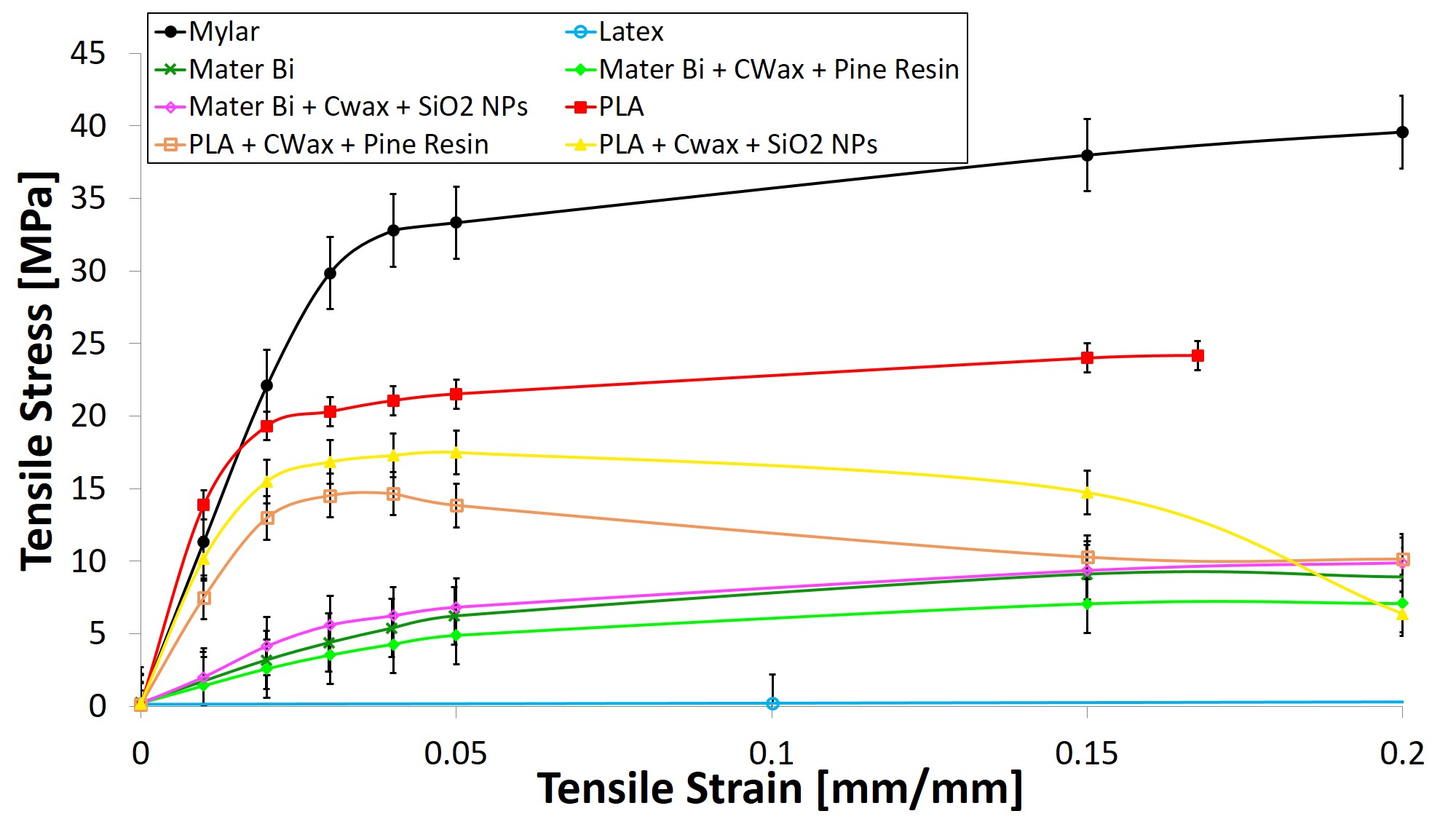}
    \caption{A zoom of the respective Tensile Stress-Strain curves for one test of each of the materials with a straining rate of 1 $mm\,min^{-1}$. This graph does not contain the average of all the results. The point of rupture is not indicated on the graph for clarity.}
    \label{fig:stress_strain}
\end{figure}

In table \ref{strength}, it can be noted that with a Young's Modulus of 1185 MPa, mylar is tougher than latex. This is because mylar contains some aluminium and the associated stress strain graph seen in fig. \ref{fig:stress_strain} is similar to those of metals. However, it has an elongation at break value of 0.8 $mm$ $mm^{-1}$ which allows it to expand slightly before rupturing. This is also the reason why it is suited for fixed-volume spheres. On the other hand, latex has the smallest Young's modulus, with a value of 4 MPa, common with most rubber materials and balloons would expand until they exploded.

As seen in table \ref{strength}, PLA with either of the two coatings is the closest in terms of toughness to mylar as they have a Young's Modulus around 1000 MPa, which could make it suitable for the radiosonde application. However, with a maximum elongation at break less than 0.1 $mm$ $mm^{-1}$, a balloon made of PLA would not be able to withstand a slight overpressure which can occur as the balloon ascends. This was experienced when trying to inflate the balloons as the PLA would easily rupture at the sealing points.  A material like Mater Bi, whose Young's Modulus is found between 100-200 MPa is more elastic than PLA and its elongation at break is around 3 $mm$ $mm^{-1}$ which means it can expand up to 300 \% of its initial size but is much lower than the 1800 \% expansion seen for latex. Hence, Mater Bi would allow for some slight overpressure without rupturing but prohibits the balloon from expanding indefinitely. This was seen when the Mater Bi balloons were inflated, as depicted in fig. \ref{fig:balloon}. Furthermore, the energy, also known as the area under the curve, is the maximum energy that the material can absorb before breaking. Mater Bi and Mylar are comparable with values around 30 $J$ $m^{-3}$, while PLA has values around 6 $J$ $m^{-3}$. What is noteworthy is that the spray coatings appear to soften both the Mater Bi  and PLA sheets. In fact, the Young's Modulus decreased for both PLA, going from around 1500 MPa to 900 MPa and Mater Bi, with values going from 160 to 120/100 MPa. Similarly, the Tensile Stress at Yield went from 18 MPa to 14/15 MPa for PLA and from 8 to 4/6 MPa for Mater Bi.

\subsection{Contact Angle}

The level of vapour pressure in clouds at a relative humidity of 100 $\%$ is either saturated or supersaturated due to the dependency of saturation vapour pressure on temperature. Hence, as the balloons' aim is to float inside clouds for as long as possible, they will be in an environment where water droplets have diameters that range from the sub micron scale up to a few hundred $\mu m$ \cite{prabhakaran2017can}. Hence, for our purposes, the highest degree of hydrophobicity, i.e. water repelling behaviour, is paramount to prohibit any of these water droplets from adhering to the surface of the balloon as it travels in and out of the clouds. This would result in an increase of weight and consequently, a decrease in the floating altitude.

\begin{figure}[h!]

\centering
\includegraphics[width=1\textwidth]{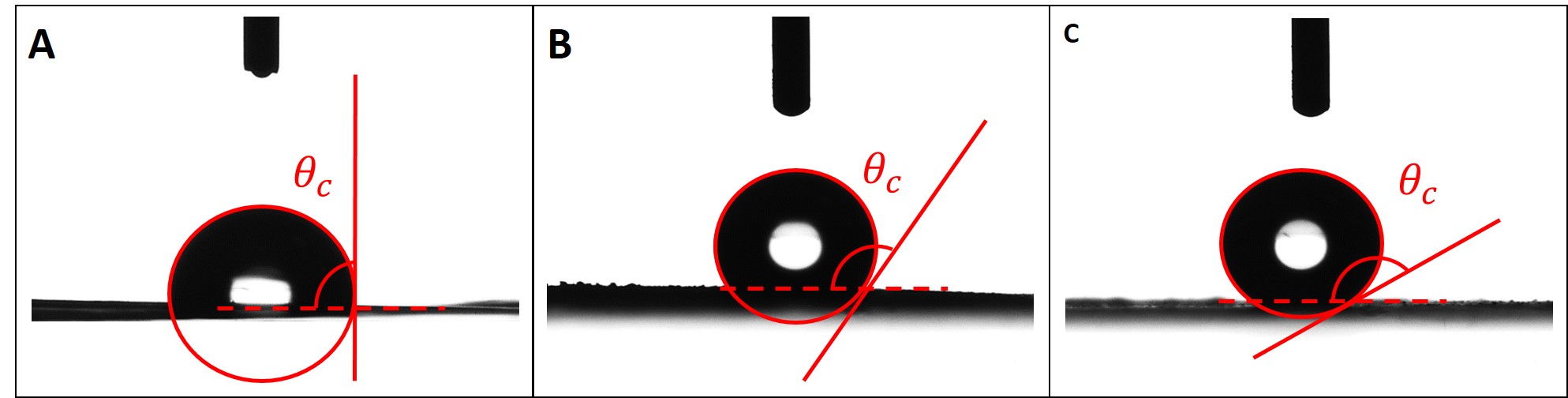}

\caption{The increase of hydrophobicity when a 5 $\mu l$ droplet is placed on A) is pristine Mater Bi, B) Mater Bi coated with carnauba wax, and C) Mater Bi coated with carnauba wax + SiO$_2$ NPs.  Note that for C), the volume of the droplet was increased, in order to obtain a picture, to 12 $\mu l$.The base diameter of the droplets were around 2 mm, see fig. \ref{fig:sem} for the visualisation of the material structure.}
\label{fig:contact_angle}

\end{figure}

\begin{table}
{\rowcolors{2}{yellow!80!white!70}{yellow!90!white!20}
\begin{tabular}{p{8cm} p{3cm}}
 \hline

\textbf{Material} & \textbf{Mean Contact Angle} [$^\circ$]\\
   
 \hline
 Latex & 79\\
 \hline
 Mylar & 95\\
 \hline
 Mater Bi & 89 \\
 \hline
 Mater Bi + Carnauba Wax & 125 \\
 \hline
 Mater Bi + Carnauba Wax + Pine Resin & 73\\
 \hline
  Mater Bi + Carnauba Wax + SiO$_2$ NPs & 140\\
 \hline
PLA & 83\\
 \hline
PLA + Carnauba Wax & 126\\
\hline
PLA + Carnauba Wax + Pine Resin & 81\\
\hline
PLA + Carnauba Wax + SiO$_2$ NPs & 136\\
\hline

\end{tabular}
}
 \caption{The average contact angle that a 5 $\mu l$ water droplet forms on the surface of tested materials. The higher the contact angle the most unfavourable to wetting the material, hence the most hydrophobic. Note that for the coatings that include SiO$_2$ NPs, the volume of the drop was 12 $\mu l$. Surfaces with values $<90^{\circ}$ are favourable to wetting and values of around 150$^{\circ}$ are obtained for superhydrophobic surfaces.}
 \label{contact}
\end{table}

With a contact angle less than 90 $^\circ$, latex is below the limit to being unfavourable to wetting and hence would not be suitable. Similarly, PLA has a contact angle closer to 80 $^\circ$ which makes it unsuitable as a material for the present application. Mylar is slightly above 90 $^\circ$ with a contact angle of 95 $^\circ$ but it is still insufficient in the context of cloud monitoring. Mater Bi is known to have some water repelling characteristics \cite{bastioli1998properties}, however, as mentioned previously, close to superhydrophobic characteristics, with a contact angle around 150  $^{\circ}$, are required. The treatment with carnauba wax was found to improve the hydrophobic properties of Mater Bi as the contact angle went from 90 $^\circ$ to 125 $^\circ$. However, for both PLA and Mater Bi, the most efficient coating was made with carnauba wax with SiO$_2$ NPs as it resulted in a coating with a contact angle of around 140 $^{\circ}$. In fact, a 5 $\mu l$ drop would not leave the dispensing needle and adhere to the surface, hence the volume of the drop that was needed to adhere onto the surface was 12 $\mu l$. The increased hydrophobicity of the carnauba wax and SiO$_2$ nanoparticles spray is in part due to the increased surface roughness which can be seen in fig. \ref{fig:sem} as well as the surface chemistry.

\subsection{He Permeability}

Another important factor that needs to be considered with balloons trying to maintain their density is the amount of helium gas, He, within the enclosure. As the He inevitably leaks out of the balloon, the altitude at which it will float will slowly decrease, hence, the key parameter to evaluate is the rate at which the material releases He \cite{murray2016helium}. The ability of the material to maintain He when fully inflated was evaluated over a time frame of 3 hours. This was chosen by considering the amount of time the initial prototypes of the probes will remain in the clouds, mostly limited by the battery. 

\begin{figure}[h!]
    \centering
    \includegraphics[width=1\textwidth]{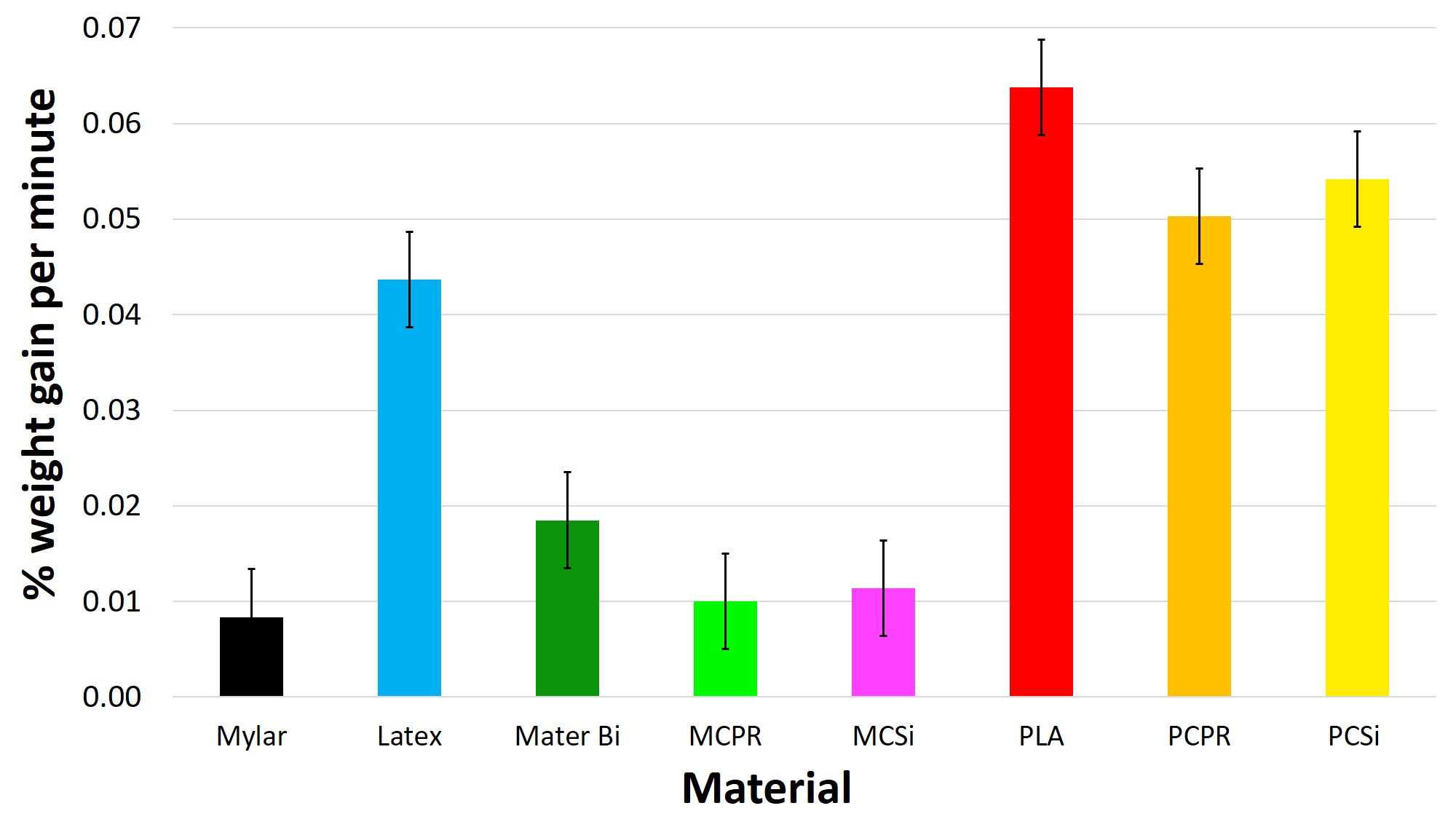}
    \caption{The percentage weight gain per minute for each material. The acronyms used, for clarity, are MCPR/PCPR is Mater Bi/PLA + carnauba wax + Pine resin coating, MCSi/ PCSi is Mater Bi/PLA + carnauba wax + SiO$_2$ NPs coating.}
    \label{fig:percentage_weight_gain}
\end{figure}

From the graph in fig. \ref{fig:percentage_weight_gain}, Mater Bi with the two carnauba wax blends coatings with a value around 0.01 $\pm$ 0.005 \% weight gain per minute seem in very close correlation with mylar whose value sits at 0.008 $\pm$ 0.005 \% which makes them promising candidates for the balloon. PLA on the other hand does not show promising results with values between 0.05 and 0.06 $\pm$ 0.04 \%. This is in part due to the increased difficulty to make a balloon with PLA as it is a rigid material and can easily form holes when inflating. Although, as before, the two coatings seem to improve the results slightly, due to the added protective layer.

If the percentage weight gain per minute of a material is in the range of 0 - 0.04 \%, it is acceptable for the present purposes. To put into perspective, for a coated Mater Bi balloon, the weight gained in 3 hours will be around 0.05 g which corresponds to a decrease in altitude of about 100 $m$. In fig.s, \ref{fig:sem} D and G, the reason for the increased He permeability for the materials with the carnauba wax and pine resin spray is made clear due to the fact that it seems to add a quasi-homogeneous polymeric layer rendering it more difficult for the He to diffuse through. This is due to the fact that permeability is also a result of dissolution of gas in the polymer, diffusion in the solid material and desorption to the exterior.

\subsection{Temperature and humidity measurements in climate chambers}

Measurements in the climate chambers of INRiM were performed to simulate the conditions that the balloons would find themselves in within clouds. Two tests were performed in the small climate chamber; the first to evaluate whether the condensation, if any, has an effect on the temperature readings and the second had the aim to understand if any water vapour was entering the balloon by measuring the humidity inside the enclosure. This is because though the electronic components of the radiosondes will be placed inside the balloon. Their accuracy depends on the stability of the conditions within the balloon hence, a material with greater inertia is required. 

By placing the sensor inside the balloon and sealing it, its inertia can be measured. Additionally, this type of measurement provides an idea of the frequency of fluctuations that the radiosondes can be sensitive to with the sensor inside the balloon. Simultaneously, the response and temperature induced stresses of the material when subject to various temperature and humidity cycles was also evaluated.

\begin{figure}
\vspace{-13mm}
    \centering
    \includegraphics[width=1\textwidth, height=6cm]{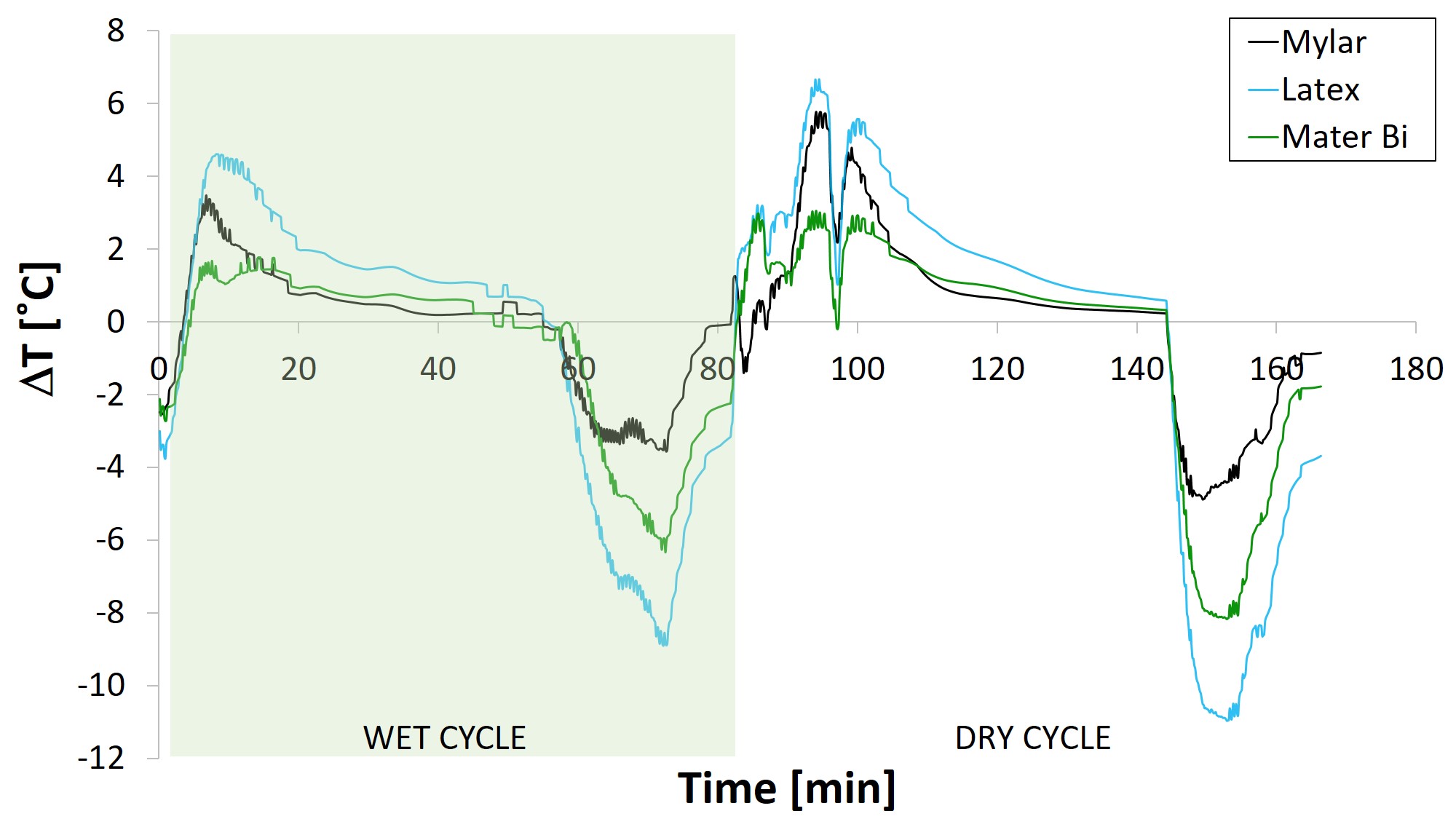}
    \includegraphics[width=1\textwidth, height=6cm]{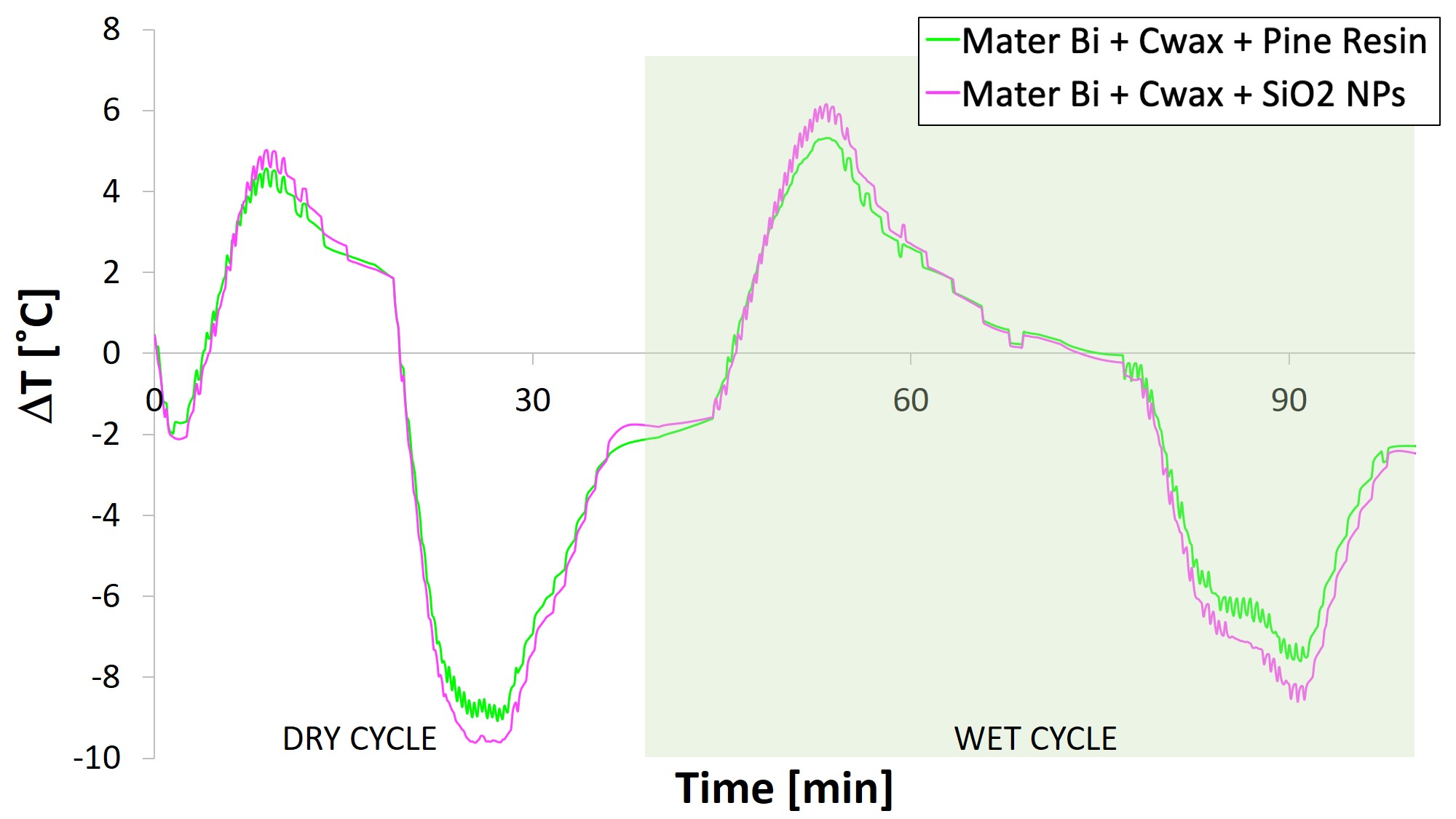}
    \includegraphics[width=1\textwidth, height=6cm]{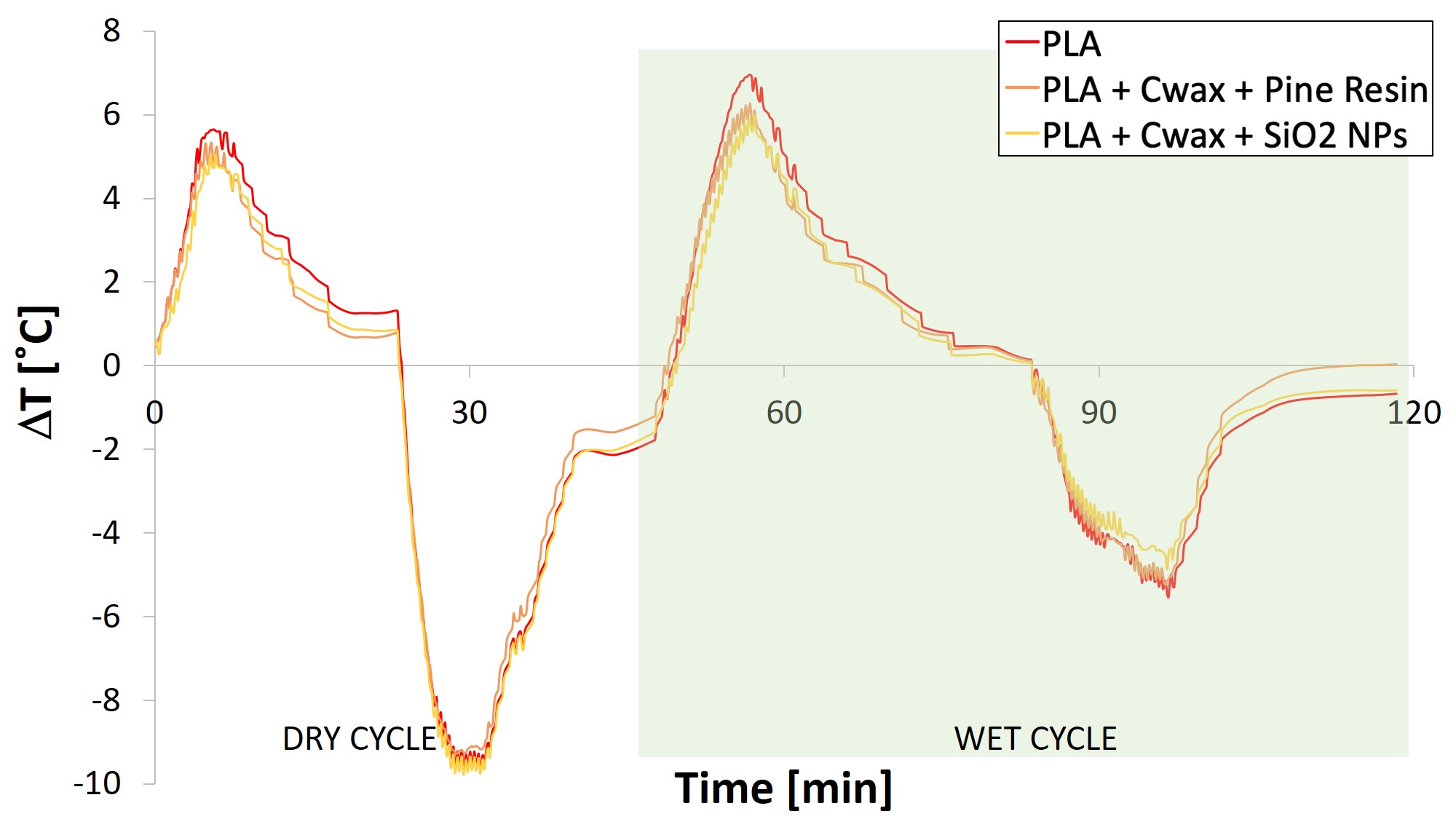}
    \caption{The response in temperature for each of the materials when subjected to a rapid decrease in temperature both during a wet cycle, indicated on the graph, where condensation is present and a dry cycle where the humidity is kept to a minimum. The dotted lines represents the moments where the temperature settings were either decreased or increased.}
    \label{fig:temperature}
\end{figure}

\begin{figure}
\vspace{-13mm}
    \centering
    \includegraphics[width=1\textwidth]{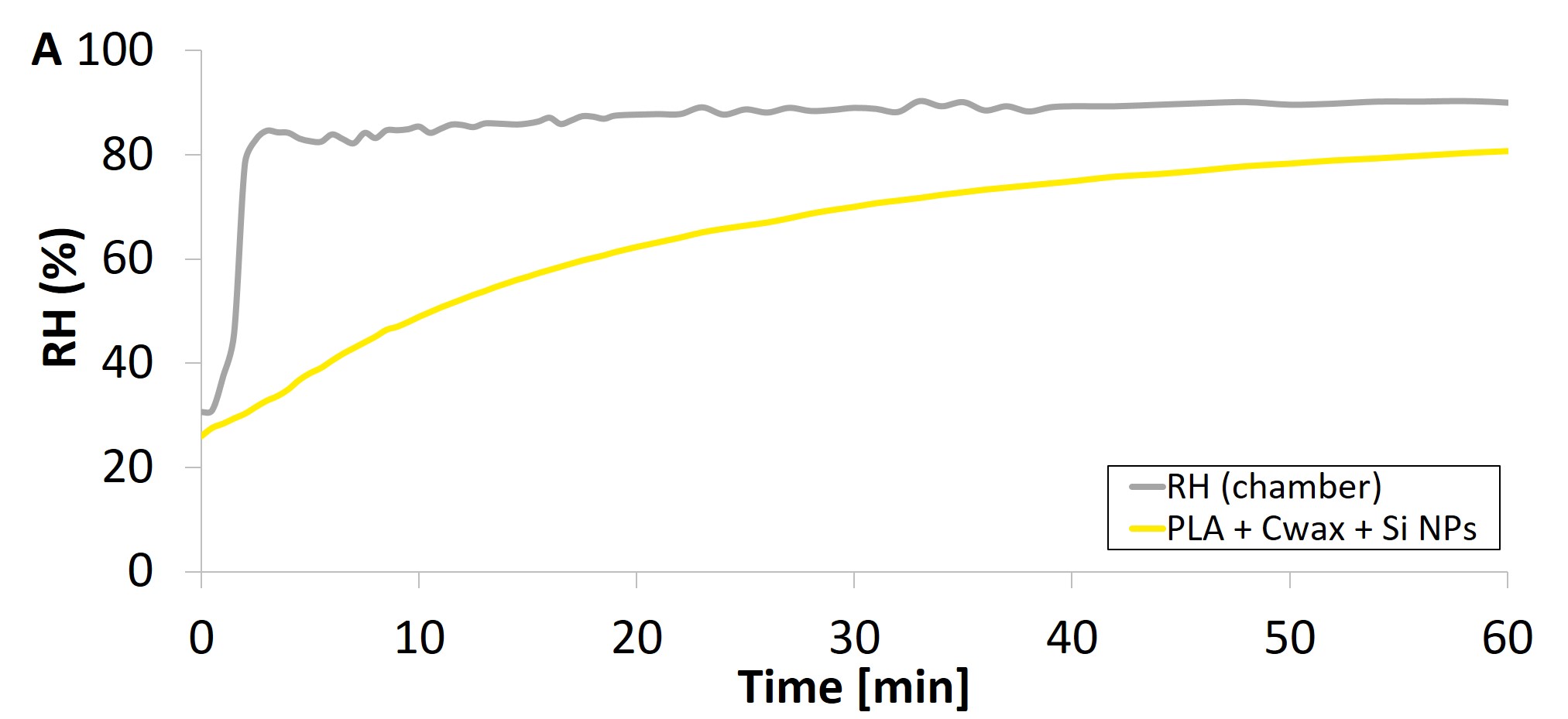}
    \includegraphics[width=1\textwidth]{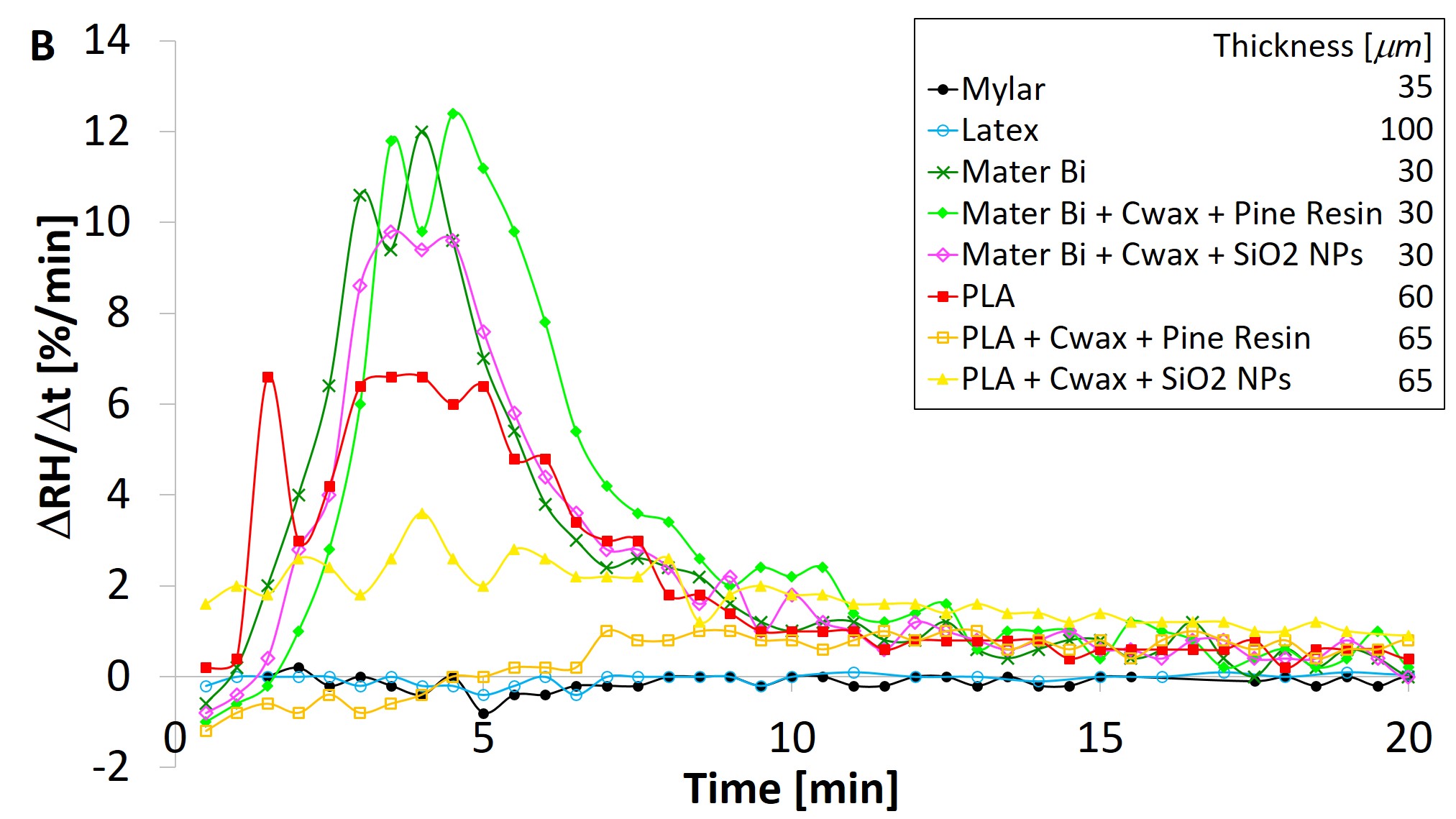}
    \caption{A) An example of the graph obtained from the humidity measurements shown for PLA + Cwax + SiO$_2$ NPs. B) The derivative of the response in humidity for each of the balloons with the corresponding thicknesses indicated in the legend. Note: after 20 minutes, all values for both balloons and chamber were more or less stable and hence not shown, for clarity.}
    \label{fig:humidity}
\end{figure}

From fig. \ref{fig:temperature}, the temperature difference between the chamber sensor and the sensor within the balloon can be seen for each different materials. Three equal tests were performed as there were only four sensors available in the chamber. All materials have a maximum difference of about 5 $^\circ$C when increasing the temperature and 10 $^\circ$C when decreasing it. Hence, they all have a very similar inertia with respect to temperature and is sufficient to keep stable working conditions for the electronics. The two materials that are perhaps less efficient are mylar and Mater Bi, with maximum values around 4 $^\circ$C and 2 $^\circ$C respectively. 

The results for the second test are presented in fig. \ref{fig:humidity}. The graph above provides an example for one material, PLA + carnauba wax + pine resin, of the response of the humidity sensor within it. The graph on the bottom is the derivative of each graph, shown for all materials for the first 20 minutes, as they then stabilise. As they had different thicknesses, their corresponding thickness are indicated in the inset box. Mylar and latex show excellent resilience to an exterior increase in humidity. The fluctuations, that can be seen in the graph, derive from the experimental uncertainty and are not meaningful. PLA + carnauba wax + pine resin also shows great promise as it has a maximum value of 2 $\% \, min^{-1}$. As predicted, the Mater Bi based materials have higher values, around 12 $\% \, min^{-1}$, as they are very thin and hence allow for a greater increase in relative humidity. To properly compare these materials, further tests will be made with thicker versions of the Mater Bi balloons. Furthermore, with a decrease ranging from 2 - 10 $\% \, min^{-1}$, the coatings seem to have a positive effect on the response velocity.

\section{Conclusions}
We have characterised two base materials, Mater Bi and PLA, with subsequent coatings and confronted them with reference balloon materials such as latex and mylar. Focusing on measurements such as contact angle, tensile strength and He permeability, we were able to fine tune some properties, being hydrophobicity, elasticity, and He retention, by means of coatings and choosing the appropriate synthesis process to obtain the exact characteristics needed for the project. 

With a Young's Modulus over 100 MPa and a maximum elongation limited to the order of 3 $mm/mm$, Mater Bi and its derivatives show promise in replacing mylar as the balloon material. This will allow some over pressure without differing too much from the chosen altitude. PLA on the other hand, doesn't allow for any elongation and would immediately burst if subject to any over pressure. This was also seen in the difference of one order of magnitude in the energy absorbed by PLA in comparison with Mater Bi or mylar. 

The hydrophobicity of each surface was tested and it was found that coating either PLA or Mater Bi with carnauba wax and silica nanoparticles resulted in a superhydrophobic surface with a contact angle around 140 $^{\circ}$. Hence, as the radiosondes are destined to travel within clouds where the humidity is 100$\%$, the balloon materials with the SiO$_2$ NPs containing coatings are the most suited. 

Furthermore, to keep the radiosondes at the chosen altitude for a maximum amount of time, the permeability to helium gas was evaluated. With a percentage weight gain less than 0.02 percentage weight gain per minute, mylar, Mater Bi and Mater Bi with both coatings are relatively impermeable to He. For a flight of 3 hours, the decrease in altitude will be of the order of 100 $m$ which is acceptable for the current purpose.

Lastly, the balloon materials were placed in climate chambers at INRiM to be tested under conditions similar to those found in clouds. The temperature and humidity response of sensors placed inside the balloons and sealed off were evaluated. All the balloon materials have an inertia with respect to temperature which is sufficient to keep the conditions of the atmosphere within the balloon stable to protect the electronics. For the measurements with humidity, PLA with either of the two coatings behaved better than the Mater Bi equivalent, however, Mater Bi with the carnauba wax and SiO$_2$ NPs showed a slower response to the other Mater Bi materials. To improve this, similar tests will be performed with thicker versions of the Mater Bi and coated Mater Bi materials. 

Therefore, Mater Bi with both the carnauba wax and pine resin, and the carnauba wax and SiO$_2$ NPs coatings are the best fit for the moment with features that still need developing. The features include the hydrophobicity of the pine resin coating and to test thicker versions of these balloons to test in the climate chambers. PLA shows good promise, it is more robust than Mater Bi, however, the synthesis and the He permeability need to be further developed and investigated to see whether it can be a true contender as balloon material.

\section{Acknowledgements}
\noindent This project has received funding from the Marie-Sklodowska Curie Actions (MSCA, ITN ETN) under the European Union's Horizon 2020 research and innovation programme. 
Grant agreement n°675675, \url{http://www.complete-h2020network.eu}.

\newpage

\section{References}









\end{document}